AL AKHAWAYN UNIVERSITY in IFRANE
School of Science & Engineering
Computer Security (CSC 3355)
Spring 2011

Project N°2:

# Authentication and Encryption in GSM and 3G/UMTS
## An Emphasis on Protocols and Algorithms

Written by *Ali Elouafiq*
Supervised by *Dr. Tajje-Eddine Rachidi*

## I- Abstract:

Mobile communication touches every aspect of our life, it become one of the major dependencies that the 21$^{st}$ Century civilizations rely on. Thereby, security is a major issue that should be targeted by communication technologies. In this paper we will target authentication and encryption in GSM and 3G/UMTS. In order to understand clearly how things work, we will start by presenting the architecture of each network, its major components, its authentication algorithms, protocols used, and KASUMI Block Cipher.

## Table of Content

## II- General Overview:
### a. Cellular Radio Networks:

- In cellular phone telecommunication another model of networks is used to insure the data exchange between the different actors of the phone services. In fact, understanding how the cellular networks work will help us understand more how security is insured in the traffic.

The Architecture:[1,2]

- The Network is structured as a *Beehive* formed of different cells, each cell is a radio base station (or a set of radio base stations) that are covering a geographic area.

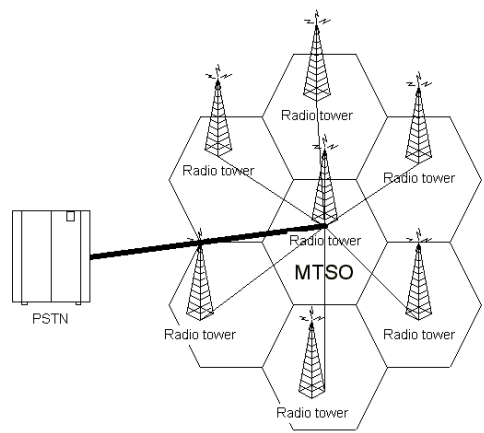

**Figure III.a.1.** Logical diagram of a cellular network [1]

- These radio base stations are connected to switching centres, mobile telephone switching office (MTSO), via fixed or microwave transmission links, that are are responsible for call processing and billing.
- These Switching centres are connected to the public switched telephone network PSTN (e.g. Internet, or other telephone networks).
- Mobile terminals (mobile phones) have a relationship with one **home network** but may be allowed to work in other **visited networks, which we call roaming,** when outside the home network coverage area, such as Going from Morocco (IAM) to France(France Telecom), or from a New York (Verizon) to Alaska (AT&T).

Location Management:[2]

- In order to transmit a call to the right person, the network should know the mobile's location, in a way that the incoming call can be routed to the correct destination. Thus, when a mobile is enters an area, it registers its current location in a **Home Location Register (HLR)** operated by the mobile's home operator.

Also, when a mobile registers in a network, information is retrieved from the HLR and stored in a **Visitor Location Register (VLR)** associated with the local switching centre.

Call establishment and handover:[2]

- When a user want to call another mobile phone, the user mobile establishes a radio connection with a nearby base station which routes the call to a switching centre. Then, before the called user receives a call, the network first tries to contact his mobile by **paging** it across its current **location area,** the mobile responds by initiating the establishment of a radio connection.
- If the mobile moves, the radio connection may be re-established with a different base station without any interruption to user communication – this is called **handover**

b. **GSM Overview[3]:**

- GSM, the *Group Special Mobile,* was a group formed by *European Conference of Post and Telecommunication Administrations* (CEPT) in 1982 to develop cellular systems for the replacement of already incompatible cellular systems in Europe. Later in 1991, when the GSM started services, its meaning was changed to *Global System for Mobile Communications* (GSM). The entire architecture of the GSM is divided into three subsystems: *Mobile Station* (MS)*, Base Station Subsystem* (BSS) and *Network Subsystem* (NSS) as shown in Figure 1.
    - The MS consists of:
        - Mobile Equipment (ME) (e.g. mobile phone)
        - Subscriber Identity Module (SIM) which stores secret information like International Mobile Subscriber Identity Module (IMSI), secret key (Ki) for authentication and other user related information (e.g. certificates).
    - The BSS, the radio network, controls the radio link and provides a radio interface for the rest of the network. It consists of two types of nodes: Base Station Controller (BSC) and Base Station (BS).
        - The BS covers a specific geographical area (hexagon) which is called a cell. Each cell comprises of many mobile stations.
        - A BSC controls several base stations by managing their radio resources, and it is connected to Mobile services Switching Center (MSC) in the third part of the network NSS also called the Core Network (CN).

NSS consists of several other databases like Visitor Location Register (VLR), HLR and Gateway MSC (GMSC) which connects the GSM network to Public Switched Telephone Network (PSTN).
- The MSC, in cooperation with HLR and VLR, provides numerous functions including registration, authentication, location updating, handovers and call routing.

- o The HLR holds administrative information of subscribers registered in the GSM network with its current location. Similarly, the VLR contains only the needed administrative information of subscribers currently located/moved to its area.
- o The Equipment Identity Register (EIR) and AuC contains list of valid mobile equipments and subscribers' authentication information respectively.

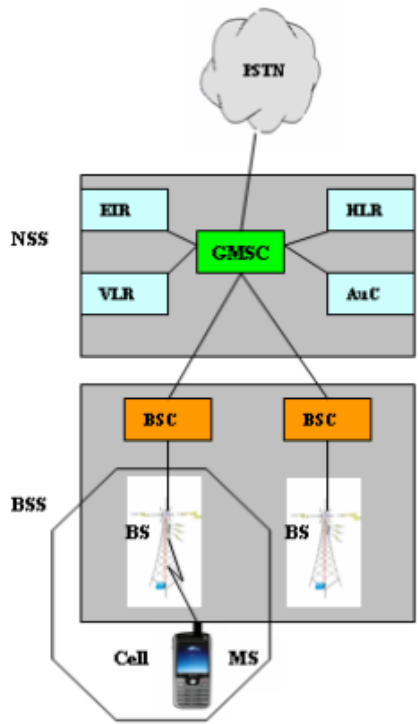

**Figure II.a.1.** Components overview of GSM [3]

c. **3G/UMTS Overview [3,2]:**

- Third generation (3G) mobile phones are characterized by higher rates of data transmission and a richer range of services. In fact, two main standards in use today the UMTS (Universal Mobile Telecommunications System) and the CDMA (Code Division Multiple Access).

- The UMTS, in fact, is the result of evolution in GSM network through GPRS. The GSM networks are capable of voice communication using Circuit Switched (CS) technique while GPRS adds Packet Switched (PS) technique through the use of some extra nodes like Serving GPRS Support Node (SGSN) and Gateway GPRS Support Node (GGSN). The UMTS, incorporating GPRS nodes and UMTS Terrestrial Radio Access Network (UTRAN), provides both circuit switched and packet switched services with enhanced multimedia applications.

### III- The Encryption and Authentication Algorithms:

All kind of networks may encounter different security threats, among these there are several known techniques such as:
- *Masquerading or ID Spoofing.*
- Unauthorized use of resources.
- Unauthorized disclosure and flow of information.
- Unauthorized alteration of resources and information.
- Repudiation of actions.
- *Denial-of-service.*

Thus telephony security is needed to provide the users of the cellular network with anonymity and privacy when making a call, to ensure the network operator bills the correct customer and to make sure that operators don't interfere with each other either accidentally or intentionally.

#### a. GSM:

- The GSM networks hold some security services for both the subscribers of the network and the operators, which are mainly reflected as subscriber's identity verification, The GSM network incorporates certain security services for operators as well as for their subscribers. It verifies subscribers' identity, keeps it secret, keeps data and signaling messages confidential and identifies the mobile equipments through their International Mobile Equipment Identity (IMEI). In the next subsections, we explain subscribers' authentication and data confidentiality as they are closely related to our topic
- Authentication:
  - The SIM card:
- The SIM card is used for the authentication in the Cellular Network, which holds different information about the machine, including:
  - IMSI, International Mobile Subscriber Identity Module.
  - Phone number.
  - Authentication key Ki.
  - Subscriber-relevant data (e.g. Text Messages and Phone Directory).
  - Security algorithms (e.g. A3).

  - The Authentication Process [2]:

- In GSM, the users are first identified and authenticated then the services are granted. The GSM authentication protocol consists of a challenge-response mechanism. The authentication is based on a secret key **Ki** which is shared between **HLR** and **MS**. After a visited MS gets a free channel by requesting BS, it makes a request for its location update to MSC through BSC. The MSC, in response, asks MS for its authentication.

- Thus in the authentication process 3 major actors are responsabile, the MS,MSC/VLR and HLR/AuC.

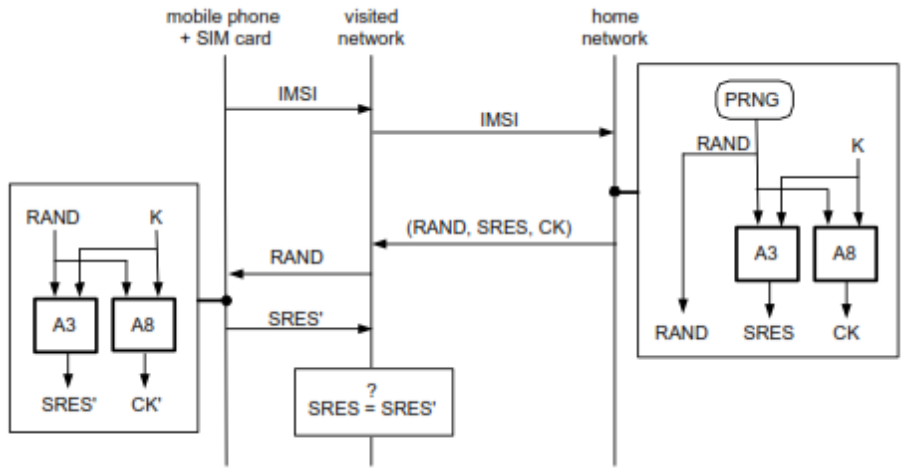

**Figure III.a.1.** The GSM authentication architecture. [7]

1- The mobile station sends its *Temporary Mobile Subscriber Identity* (TMSI) to VLR in its request for authentication.
2- The MS uses its real identity *International Mobile Subscriber Identity* (IMSI) when it is switched on for the first time but the temporary identity TMSI is used later.
3- The TMSI is used to provide *anonymity* to the user identity.
4- After getting the IMSI of the mobile station from the old VLR using TMSI the VLR sends IMSI to the corresponding HLR/AuC.
5- The HLR/AuC uses authentication algorithm (A3) and ciphering key generation algorithm (A8) to create the encryption key (Kc) and *Signed RESult* (SRES) respectively.
6- The HLR sends the triplet to VLR, including:
    1. Kc.
    2. RAND .
    3. SRES to VLR.
7- The VLR sends the RAND challenge to MS and ask to generate an SRES and send it back.
8- The mobile station creates an encryption key Kc and SRES using algorithms A3 and A8 with the inputs secret key Ki and RAND challenge.
9- It stores Kc to use it for encryption and sends SRES back to the VLR.
10- The VLR compares SRES with the one sent by HLR. If they match, the authentication succeeds otherwise it fails. [2,4,5,6]

**Note:**
- The encryption key Kc is used by both of the parties (home system and mobile station) to encrypt the data and signaling information using A5 algorithm.
- The encryption is done by mobile equipment not the SIM because SIM does not have enough power and processing capacity [2,4,5,6]

- In fact GSM systems relying on security through obscurity, which means that, the GSM algorithms will be hard to break if the attackers don't know them. Thus, the GSM specifications and protocols were kept secret away from public to be studied and analyzed by scientific community.

b. **3G/UMTS:**

- The 3G/UMTS reuse the GSM security principles by, using a removable hardware security module USIM (similar to SIM), using Radio Interface encryption, limited trust in the Visited Network, and Protection of the identity of the end user. In addition to that, 3G/UMTS added correction of weaknesses of the previous generation, which are:
    - Possible attacks from a faked base station.
    - Cipher keys and authentication data transmitted in clear between and within networks.
    - Encryption not used in some networks open to fraud.
    - Data integrity not provided.

- The UMTS, incorporating GPRS nodes and *UMTS Terrestrial Radio Access Network* (UTRAN), provides both circuit switched and packet switched services with enhanced multimedia applications.
- The circuit switched services are provided by VLR and the packet switched services are provided by SGSN. Also, the UMTS, like GSM/GPRS, deploys the concept of *Authentication Vector* (AV), unlike GSM/GPRS, the AV comprises of five components:
    - The random challenge (RAND).
    - The expected response (XRES).
    - The key for encryption (CK).
    - The integrity key (IK).
    - The authentication token (AUTN).

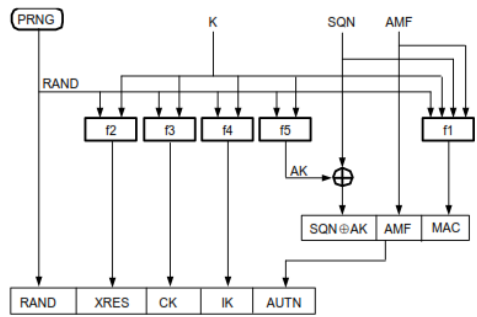

**Figure III.b.1.** Authentication vectors in 3G/UMTS

The authentication is achieved in these following steps:
1. The VLR/SGSN requests HLR/AuC for authentication.
2. The HLR/AuC computes the AV and is sent back as a response to VLR/SGSN without any encryption applied to it.
3. After the authentication is completed, the cipher key CK is used to encrypt the user data and signaling information.
4. To preserve the integrity of the important control signals, integrity key (IK) is used.

### c. Kasumi Block Cipher [8]:
#### 1- Description:
- Kasumi is a block cipher used in GSM and 3G/UMTS, In UMTS KASUMI is used in the confidentiality (*f8*) and integrity algorithms (*f9*) with names UEA1 and UIA1, respectively. In GSM KASUMI is used in the **A5/3** key stream generator.
- The algorithm uses 128-bit key and 64-bit input and output, and it based as an eight-round Feistel network.
- Using key scheduling in each round, the round function uses a round key which consists of eight 16-bit sub keys derived from the original 128-bit key using a fixed key schedule.

#### 2- Key Scheduling:

The 128-bit key $K$ is divided into eight 16-bit sub keys $K_i$:

$$K = K_1 || K_2 || K_3 || K_4 || K_5 || K_6 || K_7 || K_8$$

A key $K'$ is derived from $K$, by xoring it with *0x123456789ABCDEFFEDCBA987654321*, that is divided as well into 16- bit sub keys $K'_i$.

Round keys are either derived from the sub keys by bitwise rotation to left by a given amount and from the modified sub keys.

The round keys are as follows:

$$\begin{aligned}
KL_{i,1} &= \text{ROL}(K_i, 1) \\
KL_{i,2} &= K'_{i+2} \\
KO_{i,1} &= \text{ROL}(K_{i+1}, 5) \\
KO_{i,2} &= \text{ROL}(K_{i+5}, 8) \\
KO_{i,3} &= \text{ROL}(K_{i+6}, 13) \\
KI_{i,1} &= K'_{i+4} \\
KI_{i,2} &= K'_{i+3} \\
KI_{i,3} &= K'_{i+7}
\end{aligned}$$

#### 3- The Algorithm:

KASUMI algorithm divides the 64-bit word in two 32-bit halves, left ($L_i$) and right ($R_i$). The input word is concatenation of the left and right halves of the first round:

$$\text{input} = R_0 || L_0$$

In each round the right half is XOR'ed with the output of the round function after which the halves are swapped:

$$\begin{aligned}
L_i &= F_i(KL_i, KO_i, KI_i, L_{i-1}) \oplus R_{i-1} \\
R_i &= L_{i-1}
\end{aligned}$$

The round functions for even and odd rounds are slightly different. In each case the round function is a composition of two functions $FL_i$ and $FO_i$. For an odd round

$$F_i(K_i, L_{i-1}) = FO(KO_i, KI_i, FL(KL_i, L_{i-1}))$$

for an even round

$$F_i(K_i, L_{i-1}) = FL(KL_i, FO(KO_i, KI_i, L_{i-1})).$$

The output is the concatenation of the outputs of the last round.

$$\text{output} = R_8 \| L_8.$$

Both *FL* and *FO* functions divide the 32-bit input data to two 16-bit halves. In fact, The *FL* function is an irreversible bit manipulation while the *FO* function is an irreversible three round Feistel-like network.

### 4- FL, FO, and FI functions [8]:

**Function FL [8]**

The 32-bit input $x$ of $FL(KL,x)$ is divided to two 16-bit halves $x = l \,||\, r$. First the left half of the input $l$ is ANDed bitwise with round key $KL_{i,1}$ and rotated left by one bit. The result of that is XOR'ed to the right half of the input $r$ to get the right half of the output $r'$.

$$r' = \text{ROL}(l \wedge KL_{i,1}, 1) \oplus r$$

Then the right half of the output $r'$ is ORed bitwise with the round key $KL_{i,2}$ and rotated left by one bit. The result of that is XOR'ed to the left half of the input $l$ to get the left half of the output $l'$.

$$l' = \text{ROL}(r' \vee KL_{i,2}, 1) \oplus l$$

Output of the function is concatenation of the left and right halves $x' = l' \,||\, r'$.

**Function FO [8]**

The 32-bit input $x$ of $FO(KO,KI,x)$ is divided into two 16-bit halves $x = l_0 \,||\, r_0$.

In each of the three rounds (indexed by $j$ that takes values 1, 2, and 3) the left half is modified to get the new right half and the right half is made the left half of the next round. [8]

$$r_j = FI(KI, l_{j-1} \oplus KO_{i,j}) \oplus r_{j-1}$$
$$l_j = r_{j-1}$$

The function *FI* is an irregular Feistel-like network.[8]

**Function FI [8]**

The 16-bit input $x$ of the function $FI(KI,x)$ is divided to two halves $x = l_0 \,||\, r_0$ of which $l_0$ is 9 bits wide and $r_0$ is 7 bits wide.

Bits in the left half $l_0$ are first shuffled by 9-bit substitution box (S-box) *S9* and the result is XOR'ed with the zero-extended right half $r_0$ to get the new 9-bit right half $r_1$.[8]

$$r_1 = S9(l_0) \oplus (00||r_0)$$

Bits of the right half $r_0$ are shuffled by 7-bit S-box *S7* and the result is XOR'ed with the seven least significant bits (*LS7*) of the new right half $r_1$ to get the new 7-bit left half $l_1$.[8]

$$l_1 = S7(r_0) \oplus LS7(r_1)$$

The intermediate word $x_1 = l_1 \,||\, r_1$ is XORed with the round key KI to get $x_2 = l_2 \,||\, r_2$ of which $l_2$ is 7 bits wide and $r_2$ is 9 bits wide. [8]

$$x_2 = KI \oplus x_1$$

Bits in the right half $r_2$ are then shuffled by 9-bit S-box *S9* and the result is XOR'ed with the zero-extended left half $l_2$ to get the new 9-bit right half of the output $r_3$.[8]

$$r_3 = S9(r_2) \oplus (00||l_2)$$

Finally the bits of the left half $l_2$ are shuffled by 7-bit S-box *S7* and the result is XOR'ed with the seven least significant bits (*LS7*) of the right half of the output $r_3$ to get the 7-bit left half $l_3$ of the output.[8]

$$l_3 = S7(l_2) \oplus LS7(r_3)$$

The output is the concatenation of the final left and right halves $x' = l_3 \,||\, r_3$. [8]

### d. MM Protocol [9, 10]:

In UMTS and GSM, authentication is processed by the Mobility Management layer. The format of the header is as follow

| 8 | 7 | 6 | 5 | 4 | 3 | 2 | 1 | Octet |
|---|---|---|---|---|---|---|---|---|
| Protocol discriminator | | | | Skip indicator | | | | 1 |
| Message type | | | | | | | | 2 |
| Information elements | | | | | | | | 3-n |
| MM header structure | | | | | | | | |

**Protocol discriminator**
0101 identifies the MM protocol.

**Skip indicator**
The value of this field is 0000.

**Message Type:**

The message type defines the function and format of each MM message. The 8th is reserved for possible future use as an extension bit, the $7^{th}$ bit is reserved for the send sequence number in messages sent from the mobile station. MM message types that are related to authentication:

0x01: xxxx Security messages:
0001 AUTHENTICATION REJECT
0010 AUTHENTICATION REQUEST
0100 AUTHENTICATION RESPONSE
1000 IDENTITY REQUEST
1001 IDENTITY RESPONSE
1010 TMSI REALLOCATION COMMAND
1011 TMSI REALLOCATION COMPLETE

## IV- Conclusion:

- Wireless communication is attractive to both users and service providers. However when the technology popularity increases security problems of confidentiality, integrity, and authentication are also increasing. In order to understand in depth the issues that may arise, we've covered the whole network architecture, the authentication process, the encryption algorithms, and the protocol. In fact, many offensive security works are targeting this field (for more materials check Blackhat.com website), as well as formal security research such as asymmetric cryptography.

### V- Bibliography: